\documentclass[amssymb, prx, 10pt, aps,notitlepage]{revtex4-1}
\pdfoutput=1

\usepackage{latexsym,amsmath,amsfonts,amssymb,bm,empheq}   
\usepackage{graphicx}
\usepackage{booktabs}
\usepackage{supertabular}
\usepackage[dvipsnames]{xcolor}
\usepackage{subfigure}
\usepackage{hyperref}
\usepackage{float}

\definecolor{nblue}{RGB}{28,130,185}
\definecolor{cgreen}{RGB}{76,153,0}
\definecolor{myorange}{RGB}{245,156,74}

\usepackage{hyperref}
\hypersetup{
  colorlinks=true,
  citecolor=magenta,
  urlcolor=-myorange
}

\begin{document}

\title{Critical behaviour at the onset of synchronization in a neuronal model}

\author{Amin Safaeesirat}
\affiliation{Physics department, Sharif University of Technology, P.O. Box 11155-9161, Tehran, Iran}
\author{Saman Moghimi-Araghi}
\affiliation{Physics department, Sharif University of Technology, P.O. Box 11155-9161, Tehran, Iran}


\begin{abstract}
The presence of both critical behaviour and oscillating patterns in brain dynamics is a very interesting issue. In this paper, we consider a model for a neuron population where each neuron is modeled by an over-damped rotator. We find that in the space of external parameters, there exist some regions that system shows synchronization. Interestingly, just at the transition point, the avalanche statistics show a power-law behaviour. Also, in the case of small systems, the (partially) synchronization and power-law behavior can happen at the same time.
\end{abstract}

\maketitle


\section{Introduction}

Scale-free and critical behavior has long been observed in animal and human brain experiments, both in vitro and in vivo \cite{Beggs,Peterman,Gireesh,Hahn,Priesemann,Pasq}. Neuronal avalanches, the bursts of activity and spiking of neurons that spread in large areas of the brain network, exhibit power-law forms in their size and duration distributions. Also, other data like the data of the human electroencephalography (EEG) show similar patterns in the background activity of the brain \cite{Freeman,Shriki}. 

The criticality of neuronal systems is supported by the existence of power-law behavior and data collapse. Being at criticality is thought to play a crucial role in some of brain fundamental properties like optimal response and storage and transfer of information \cite{Shew,Kinouchi,ShewPlenz,Gautam,Legenstein}. As there is no need to tune any parameter to arrive at criticality, it was suggested that the brain is an example of Self-Organized Critical (SOC) systems\cite{Hesse}. The idea of self-Organized criticality was introduced by Bak, Tang, and Wiesenfeld \cite{BTW}, where using a simple sandpile model shows that some complex systems can organize themselves towards their critical point and therefore, there is no need to regulate external parameters. A range of models was introduced to show that self-organized criticality is plausible in the neural systems \cite{Bornholdt,De Archangelis,Levina,Millmann,Meisel,Tetzlaff,Kossio}. However, being at criticality does not necessarily imply that the system arrives at criticality through a self-organizing mechanism. Rather, it has been shown that neural systems may not be an example of SOC systems \cite{Bonachela}. Also, for the brain network, which has a highly hierarchical and modular structure, the critical points are generalized to critical regions, similar to that of Griffiths phases \cite{Moretti,Moosavi} , and there may be no need to fine-tune a parameter. Actually, it seems that a long term evolutionary mechanism would be enough to explain why the brain works at criticality.

Therefore, some efforts have been made to find how power law can happen in models of the neuronal system even you need to tune some external parameters\cite{Zare}. Yet it remains to understand what is the phase transition that determines the critical state. One of the most promising candidates is the synchronization transition. It is found that the brain is more or less maintained on the transition point to synchronization \cite{Gautam,Poil,Dalla}, and the criticality in avalanche statistics can happen at this transition point\cite{Fontenele,Montakhab}. Also, a general framework based on Landau-Ginzburg theory has been proposed to describe the dynamics of cortex and it is found that at the edge of synchronization scale-free avalanches appear \cite{Munoz}. In similar studies,  it is found that the bifurcation that leads to successive firings of a neuron might have effect on the overall dynamics \cite{}. 

In this paper, we study a network of neurons modeled by over-damped rotators to mimic type I neurons, rather than the usual integrate-and-fire model that more or less mimics type II neurons' behaviour. Therefore, our setup is similar to that of the Kuramoto model\cite{Kuramoto} and hence each oscillator is identified by a single-phase variable. This helps us to study the synchronization of the system more easily. Of course, the dynamics, especially the way the two neighbours affect each other is very different from the Kuramoto model and therefore it has completely different features. We explore the phase space of external parameters and find the transition lines of synchronization. The main parameters we have explored are the external stimulation power, the relative strength of inhibitory neurons to excitatory ones and the axonal time delay. We analyze the system's avalanche size statistics at different points of the phase space of the parameters and interestingly observe that right at the transition the avalanche size and period statistics obey power-law.



\section{Models and Methods: The single neuron model}
There are lots of models for single neuron dynamics. However, for simulating relatively large networks, simple models should be chosen. A simple model only keeps the main characteristics of the system and throws away other details. For a neuron, this main characteristic may be stated as follows: If a relatively small stimulation is applied to a neuron, It doesn't react considerably. However, a neuron that receives a slowly varying current throws off a fairly regular set of spikes and in this situation, it may be regarded as an oscillator.  In the language of dynamical systems, when a neuron is sufficiently stimulated, the stable fixed point is lost and its movement in phase space can be regarded as a limit cycle, at least in short times.  Therefore, for the single neuron model, it is good to consider a simple model that has such a cycle when it is stimulated. The simplest model might be the well-known leaky integrate and fire model (LIF); however, the LIF model. As we will discuss later, this model more or less mimics the behavior of Type II neurons, that is,  there is a "discontinuity" in the frequency of successive firings of the modeled neuron, or its activity, at the onset activity. In fact the activity is continuous but with a logarithmic singularity which in some sense is equivalent to discontinuity. In contrast, Type I neurons are characterized mainly by the appearance of oscillations with arbitrarily low frequency as the current is injected. In our study, we prefer to use a simple model that behaves more closely to a type I neuron. 

A suitable model for type I neurons is given by the equation
\begin{equation}\label{SingleNeuron}
    \frac{d\phi }{dt}=I-\sin\phi,
\end{equation}
where $I$ is the external stimulation and $\phi$ is the dynamical variable. The model has been considered before as a simple model to simulate the dynamics of neurons and is closely related to quadratic non-linear integrate and fire model\cite{Ermentrout-1,Ermentrout-2,Ermen}. 

The equation (\ref{SingleNeuron}) can be interpreted as an over-damped rotator in the following way: Consider a rod pendulum of mass $m$ and length $l$ driven by a constant torque in a viscous medium. The equation of motion of such pendulum is given by:
\begin{equation}\label{Rotatingrod}
ml^2\frac{ d^2 \phi }{d t ^2} + b \frac{d \phi}{d t } + mgl \sin{\phi} = T
\end{equation}
where $\phi$ is the angle of the rotating rod with the downward vertical, $b$ is the viscous damping constant, $g$ is the acceleration due to gravity, and $T$ is the external torque. 

Equation (\ref{Rotatingrod}) is a second-order system; however, in the limit of very large $b$'s, where the rotator is over-damped, it may be approximated with a first-order system. Neglecting the inertia term $m l^2 \Ddot{\phi}$ and non-dimensionalizing the resulted equation, one arrives at equation (\ref{SingleNeuron})
\cite{Strogatz}. 

Although equation (\ref{Rotatingrod}) has a very rich dynamics \cite{Strogatz}, the over-damped limit is sufficient for modeling single neuron dynamics. It shows a saddle-node bifurcation at $I=1$, when $I<1$ there exist two distinct fixed points, one of which is stable and the other is unstable. If $I>1$, there will be no fixed points and the rotator will overturn continually. The rest state of the neuron corresponds with the stable fixed point and when $I>0$ and the rotator overturns continually, the neuron fires repeatedly.

It is more convenient to take $\theta=\phi-\pi/2$ and rewrite Eq. \ref{SingleNeuron} as
\begin{equation}\label{SingleNeuront}
    \frac{d\theta }{dt}=I-\cos\theta.
\end{equation}



\begin{figure}
  \includegraphics[width=8cm]{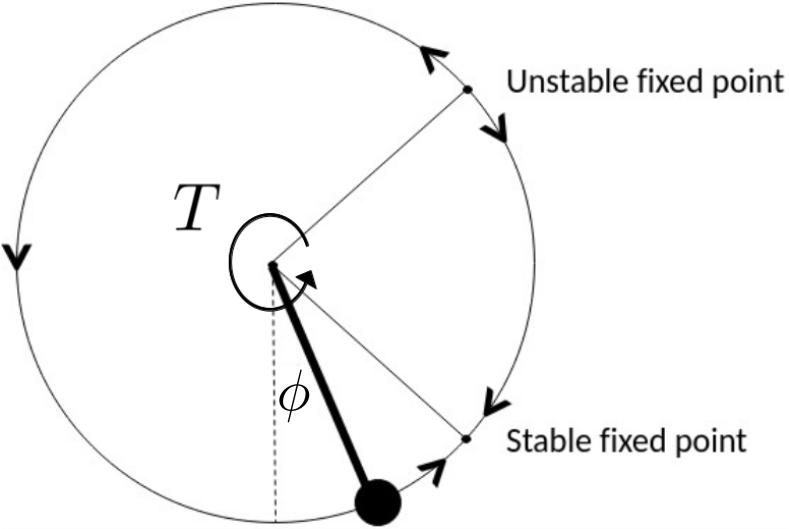}
  \caption{Two fixed points of the system for $I<1$ and $\alpha>>1$. Here $T$ is the external torque exerted on the system from outside (take a look at Eq. \ref{Rotatingrod}). The system will stand in its stable fixed point if there is no external stimulation. If the stimulation is enough and the system proceeds its unstable fixed point, it will return to the stable fixed point counterclockwise. We define this as a "spike" in our system.   }
  \label{Fixed points}
\end{figure}

 Consider the case $I<1$. The two equilibrium points are shown in figure \ref{Fixed points}. 
When the system is at rest, it stands at the stable fixed point. If the rotator is stimulated slightly, it returns back to the stable fixed point. However, if the stimulation is large enough to take the rotator beyond the unstable fixed point, or the threshold, it will rotate a complete circle and then arrives at the stable fixed point. We call this complete rotation of the rotator a spike. Of course, usually, it is considered that the "neuron" spikes when $\theta=\pi$, that is, it produces an action potential right at that moment. This model can be mapped to quadratic integrate and fire neuron by the transformation $u=\tan(\theta/2)$\cite{Ermen}.  
It is easy to show that Eq. \ref{SingleNeuront} models type I neurons. If the current applied to the neuron exceeds the threshold value $I=1$, it spikes repeatedly. The inter spike interval (ISI) is determined via
\begin{equation}
\rm{ISI}= \int_{t_f}^{t_f+{\rm{ISI}}} dt = \int_{\theta_f}^{\theta_f +2\pi} \frac{2\pi d\theta}{I-\cos{\theta}}= \frac{2\pi}{\sqrt{I^2-1}}
\label{forth-equation}
\end{equation}
where $t_f$ is the spike time and $\theta_f$ is the spike angle. Therefore, the gain function is 
\begin{equation}
f(I)= \frac{1}{\rm ISI}=\frac{\sqrt{I^2-1}}{2\pi} .
\label{Gainfunction}
\end{equation}
and hence it grows as $\sqrt{I^2-1}$ for currents slightly above the threshold current. Note that the gain function behaves as $(I-1)^\mu$ with $\mu>0$ and therefore the function is continuous. For $\mu<0$, the function is clearly this continuous and for the case of $\mu=0$ it is usually regarded that there is jump in the function and is therefore discontinuous.  In the case of the leaky integrate-and-fire model with the dynamics given by $\dot{v}=I-v$ and the threshold potential $v_{\rm Th.}=1$, the gain function is proportional to $-1/\log(I-1)\simeq (I-1)^0$, which resembles a discontinuity, although the function is actually continuous. Therefore,  to  mimic the type I neurons, we prefer to choose the overdamped rotator as our single-neuron model.

\section{Models and Methods: the network of neurons}


We have considered a network composed of $N$ neurons modeled by the over-damped rotator where $80\%$ of them are randomly chosen to be excitatory neurons, and the rest are inhibitory ones. Each neuron, regardless of being excitatory or inhibitory, receives a constant number of internal connections, that is from a fixed number of neurons within the network. We denote this number by $C$ and write $C=C_E+C_I$ where $C_E$ and $C_I$ are the number of connections from excitatory and inhibitory neurons respectively. We have considered the ratio of inhibitory inputs to excitatory ones is the same as the population ratio, which is $C_I/C_E =1/4$.  The neurons also receive $C_{ext}$ connections from external neurons. $C_{ext}$ is taken to be equal to the number of excitatory connections from the network ($C_{ext}=C_E$). The external neurons fire randomly so that the intervals between two successive firings have a Poisson distribution with the average  $T_{ext}=1/\nu_{ext}$. 
These external neurons effectively play the role of external (noisy) current, therefore we set $I=0$ in the dynamics of over-damped rotators (Eq. \ref{SingleNeuront}) and control the external current through tuning $\nu_{ext}$. 

When the rotator $i$ arrives at $\theta=\pi$, it fires and hence changes the phase of its neighbour $j$ by the amount $J_{ij}$. For presynaptic excitatory neurons $J_{ij}=J$ and for presynaptic inhibitory ones $J_{ij}=-gJ$. The same strength $J$ is considered for external (excitatory) neurons. Also, we assume that if the neuron $i$ fires at time $t$, the spike arrives at the neuron $j$ at $t+D$, that is, an axonal time delay introduced to the system. 

Now we can write the dynamics of the neuron $i$ in the following form 
\begin{equation}
\frac{\partial\theta_i}{\partial t}=-\cos\theta_i+ \sum_j J_{ij}\sum_k \delta(t-t_j^k -D),
\label{equation1}
\end{equation} 
where the first summation, is on presynaptic neurons connected to the neuron $i$, and the second summation is on the spikes of the neuron $j$. In fact, the delta function $\delta(t-t_j^k -D)$ represents the $k$th spike from the neuron $j$ arriving at $t= t_j^k+D$.

Usually, the models describing networks of neurons have a handful of parameters and are difficult to analyze completely. In our model, the parameters $\nu_{ext}, g, D, J$, and $N$ are the five adjustable parameters. The role of each parameter is obvious; $\nu_{ext}$ controls the strength of external drive, $g$ controls the relative strength of inhibition to excitation within the network, $D$ is the axonal time delay, $J$ controls the strength of the synapses and $N$ is simply the total number of neurons. To make the parameter space smaller, we have fixed $J=0.015$ and $C=100$. We try to investigate the dynamical characteristics of the system by exploring considerable parts of the remaining  3-dimensional space.

\section{Synchronization of the neurons}

We have simulated populations of the neurons to find out if synchronization happens throughout the system or not. It has been observed that in such neuron populations synchronization might happen\cite{Brunell,Montakhab,Shahbazi,Luccioli}. For example, in \cite{Brunell} a network similar to ours is studied and it has been observed that in the presence of delay, one can find some regions where the system shows synchronization. In his paper, Brunell has studied its network for different values of external drive and the relative strength of inhibition to excitation in the system and has found regions in this space where synchronization happens. Also in \cite{Valizade17}, using a Kuramoto-like model it has been shown that in presence of delay a transition from a asynchrony to synchrony happens when the level of input to the network is changed. Note that although we have rotators as our single neuron model, it differs from Kuramoto model, both in single agent dynamics and and in interactions between two adjacent neurons: it has a preferred angle $\theta_0$ (corresponding to rest state of the neurons) and, when a neuron fires, it kicks the post-synaptic neuron. In Kuramoto model, there is no preferred angle and also the interactions comes from a term proportional to sine of the angle difference of the two neuron. Quite recently, a model has been proposed in which all the above terms are present: the intrinsic angular velocity of the rotator and the sine terms of the  Kuramoto and the preferred rest point and the kicks of our model \cite{Monuz20}. Within this model both synchronous and asynchronous phases have been observed and  a rich phase diagram has been obtained. Also in some transition points scaling laws in avalanche statistics is found.    

\begin{figure}[t]
  \includegraphics[width=9cm]{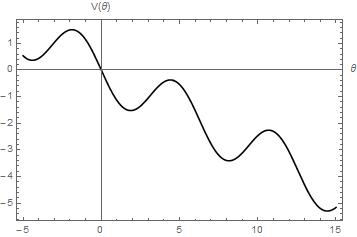}
  \caption{The plot illustrates the effective potential for the rotator. The rotator tends to stay at the local minimums of the function where $\theta= 2k\pi - \cos^{-1}I$ in which k is an integer. Due to the outside stimulation, over time, the rotator will reach to the local maximum and consequently spikes and goes to the next relative minimum. The plot is sketched for $I=0.3$}
  \label{potential-fig}
\end{figure} 

To distinguish the synchron phase from asynchron one, we need a suitable order parameter which vanishes when the system is not synchronous and takes non-zero value otherwise. In the context of Kuramoto-type rotators \cite{Kuramoto}, the order parameter is defined as  $r=|(1/N)\sum_{k=1}^N e^{i\theta_K}|$, where $\theta_k$ is the phase of $k$'th rotator. If all the rotators are in phase, $r$ will be unity and if the rotators are all independent, it would be of the order of $1/\sqrt{N}$, which becomes negligible for sufficiently large systems. Hopefully, we have phases as our dynamical variables and at the first sight, one thinks of the same order parameter. However, as we discuss below, we find this definition inappropriate for our model. Our single-neuron model is given by Eq. \ref{SingleNeuront}. To have a better insight let us interpret the same equation as a Langevin-type equation:
\begin{equation}
     \frac{d\theta_i }{dt}=I-\cos\theta_i+\eta_i=-\frac{\partial V(\theta_i)}{\partial \theta_i}+\eta_i,
\end{equation}
where we have collected the effect of network spikes and external spikes in the noise term $\eta_i$ and 
have defined the potential $V$ as
\begin{equation}
    V(\theta)=-I\theta +\sin\theta.
\end{equation}
This potential function is sketched schematically in figure \ref{potential-fig} for $I=0.3$.

The rotator tries to reach the local minimum, which is identified as the rest state of a neuron. However,  because of the noise term which is resulted from stimulations received by the neurons, it may go up-hill
and then fall off to the next minimum (and as a result fire). Therefore, it is very probable to find the rotator at the point $\cos\theta=I$, or equivalently find the neuron at rest. In such a situation, even if the firings of the neurons are irregular and asynchron, the order parameter can take a non-vanishing value. In fact, if the ratio of the neurons at rest to the total number of neurons is $\mu$, the order parameter will be $\mu+O(1/\sqrt{N})$. Therefore, especially for small external stimulation, this order parameter is useless and it doesn't show synchronize "activity".

If we want to see if the "activity" is synchronous, it is better to consider just the neurons that are firing. In the case of rotators, we may consider only the rotators with the phase parameter in the interval  $[\pi/2,3\pi/2]$. We will call these rotators, the spiking rotators. However, within this interval, the real part of $e^{i\theta}$ is always negative and it's average will be never zero. On the other hand, within the same interval, the imaginary part of $e^{i\theta}$, ( $\sin\theta$) ranges from -1 to 1. Consider the quantity $r_1=(1/N_a)\sum_k \sin\theta_k$, where $N_a$ is the number of spiking rotators and the summation is just over the set of spiking rotators. If the network activity is irregular, then $r_1$ will be of the order of $1/\sqrt{N_a}$, however if all the rotators spike synchronously, that is $\theta_k(t)\simeq f(t)$ for all spiking rotators, then $r_1= \sin(f(t))$. This function fluctuates in the interval $[-1,1]$, but we need a just number to distinguish the synchronous phase from asynchronous one. This leads us to the last step to define the proper order parameter as

\begin{equation}
m=\langle (\frac{1}{N_a}\sum_{i=1}^{N_a}\sin{\theta_k})^2\rangle,
\label{order_parameter}
\end{equation}
where $\langle .\rangle$ means time averaging over a period comparable with the spike period. In this way, if the network activity is irregular, $m=O(1/N_a)$ and in the case of synchronous activity $m=O(1)$. Just we have to check $N_a$ is not very small, and as we will see in our model, in all the regions we are interested in, the activity is large enough.

\begin{figure*}[ht]
\centering
\begin{tabular}{cc}
    \includegraphics[width=0.47\linewidth]{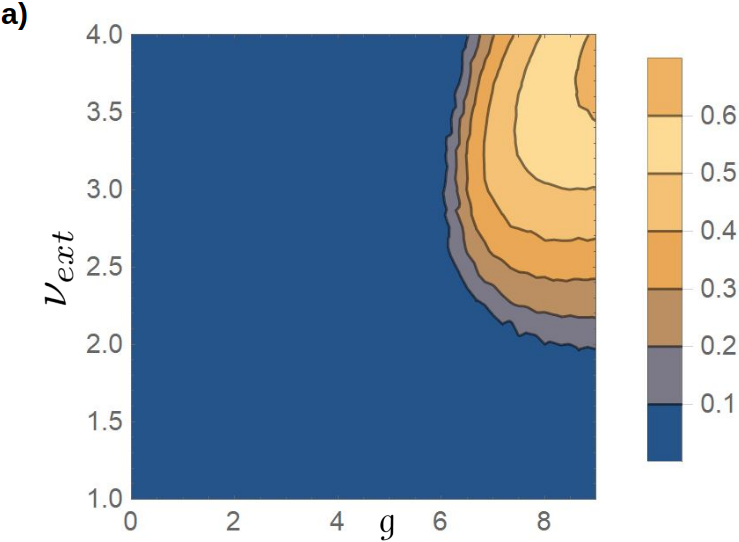}&
    \includegraphics[width=0.47\linewidth]{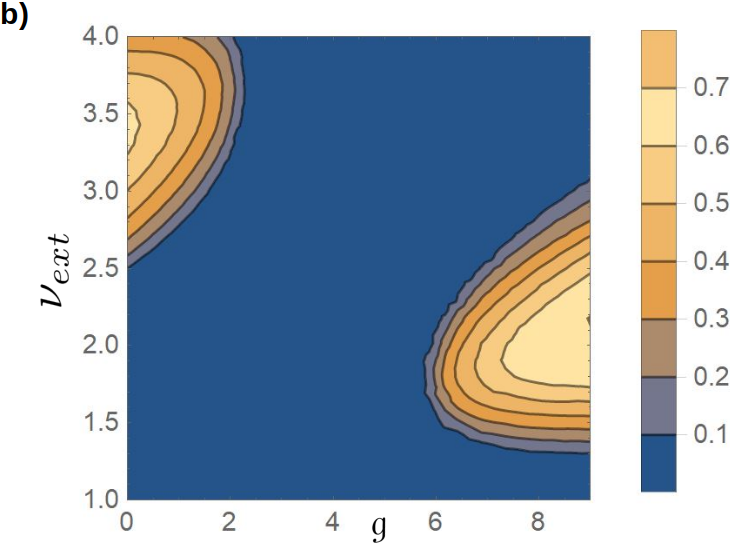}\\
    [2\tabcolsep]
    \includegraphics[width=0.47\linewidth]{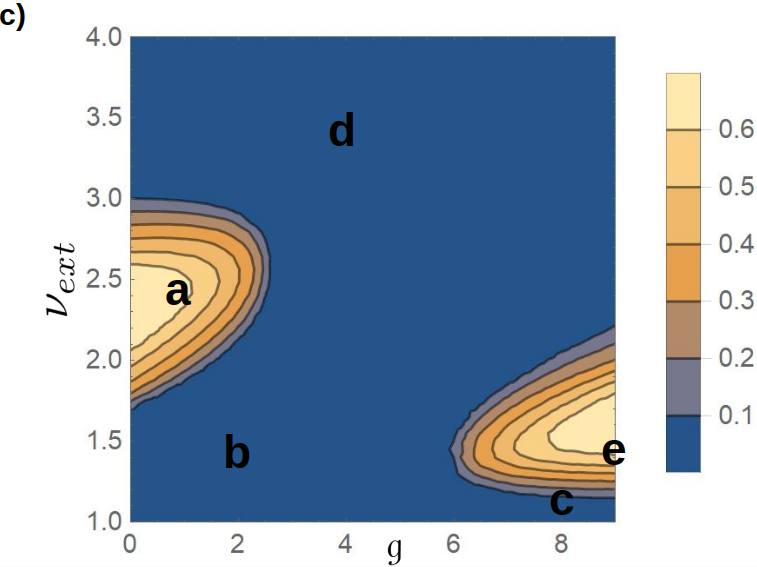}&
    \includegraphics[width=0.47\linewidth]{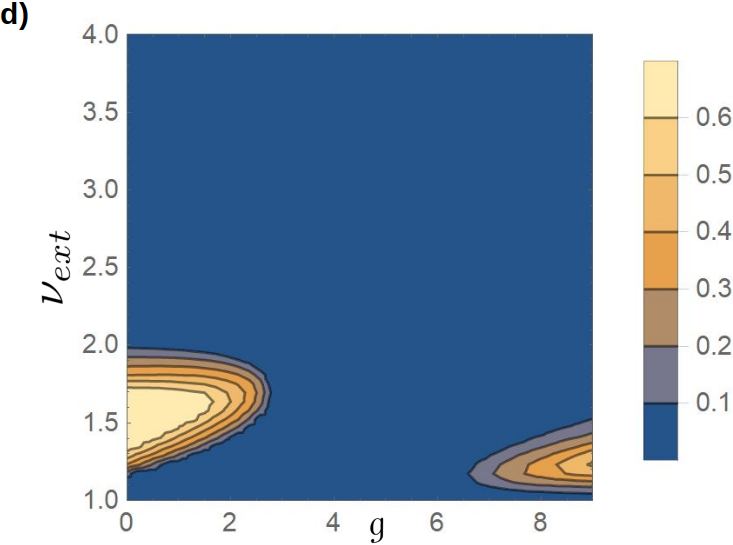}
\end{tabular}
\caption{The contour plots show the order parameter of systems of 4000 neurons for different values of $g$ (inhibitory strength), $\nu_{ext}$ (average external stimulation frequency), and delay (Eq. \ref{order_parameter}). The brighter areas in the plots are the synchronous ones. The color varies between blue and pink so that the higher the order parameter, the color is closer to pink. Not only is the delay necessary for the existence of the synchronization but also the different values of delay can change the location of the synchronous regions. As the delay increases, they are pushed to the lower $\nu_{ext}$ and dwindling. a) $D=0.5$, b) $D=1$,  c) $D=1.5$ and d) $D=2.5$. The resolution of the graphs are of $\Delta g= 0.15$ and $\Delta \nu_{\rm ext}= 0.06$. The points shown in plot "c", are selected to be representatives of different states for this system. The activity pattern and correspondent histogram of size and period of avalanches for these points are shown in Fig. \ref{activity plots} and Fig. \ref{avalanche-general} respectively.}
\label{phasediagram}
\end{figure*}

In simulations, we have considered several samples with different sizes ranging from 1000 neurons to 16000 ones. Also, we have considered several random configurations of links with the above-mentioned statistics. As for the initial condition, we have assigned a random phase between $0$ and $2\pi$ to each rotator and have numerically integrated the differential equations. The time steps were set to $\Delta t=0.01$. After the system arrives at the steady-state, which happens typically no more than 20000 temporal steps, we begin our actual measurements.

A central quantity of a neuron population is its activity. The activity time series $A(t,\delta t)$ is defined as the number of spikes in the network between $t$ and $t+\delta t$. The parameter $\delta t$ should be taken much smaller than the average period of a single neuron successive firings, but not too small that only single firings are found within each interval. We have taken $\delta t$ of the order of $\Delta t$ which seems appropriate for our systems, therefore we fix $\delta t=\Delta t$ and will simply denote $A(t,\delta t)$ by $A(t)$.

Within the same networks, we have calculated the order parameter $m$ (Eq. \ref{order_parameter}) for different values of $\nu_{ext}$, the external stimulation strength, $g$ relative number of inhibitory neurons to excitatory ones ,and the delay parameter $D$.  Figure \ref{phasediagram} shows the order parameter in $\nu_{ext}$-$g$ plane for four different values of $D$. We have considered $\nu_{ext}>1$ where the network is active.  In the simulation we have set $N=4000$ and $J=0.015$ and $D$ takes the values 0.5, 1, 1.5 and 2.5 in sub-figures (a) to (d) respectively. For all values of $D$, there are two distinct regions that the order parameter takes non-zero values. These regions are pushed towards the line $\nu_{ext}=1$ as $D$ is increased. In the case $D=0.5$, only one of the two regions is seen in the graph as the other is at much higher values of $\nu_{ext}$. We have also checked the case of $D=0$ and within a larger area (up to $\nu_ext=60$) and observed no synchronous phases. In other words, the presence of the delay in the system plays a crucial role in the properties of this collective behaviour. For a larger delay, smaller external stimulation is needed to arrive at synchronization. We will denote the left region by L-Synch. and the right one by R-Synch. In the L-Synch. region, the excitatory neurons dominate and in R-Synch. region the inhibitory neurons dominate. 

\begin{figure}[t]
 \centering
   \includegraphics[width=0.48\textwidth]{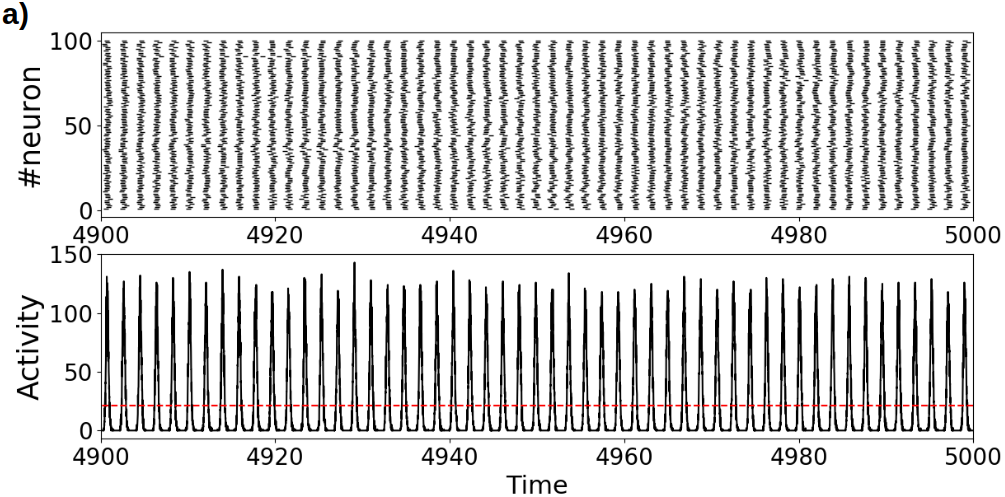}
   \includegraphics[width=0.48\textwidth]{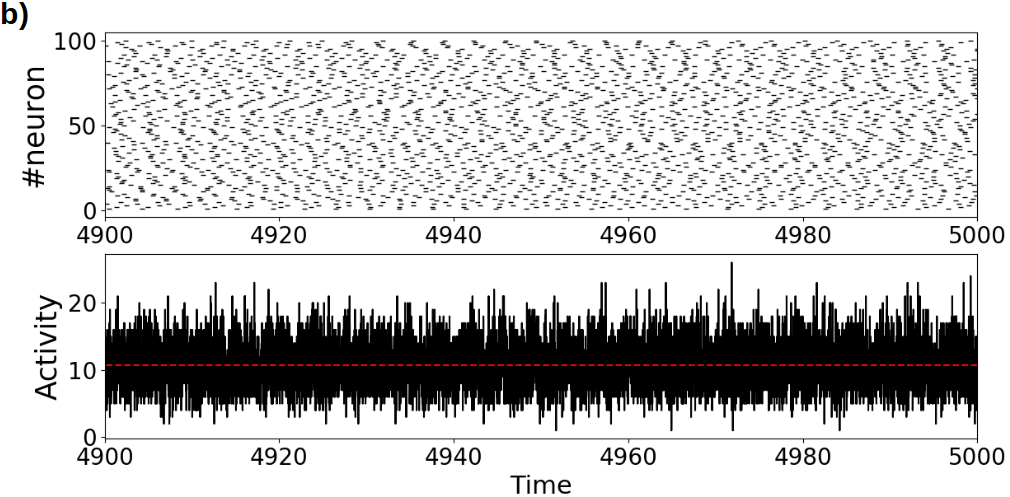}
   \includegraphics[width=0.48\textwidth]{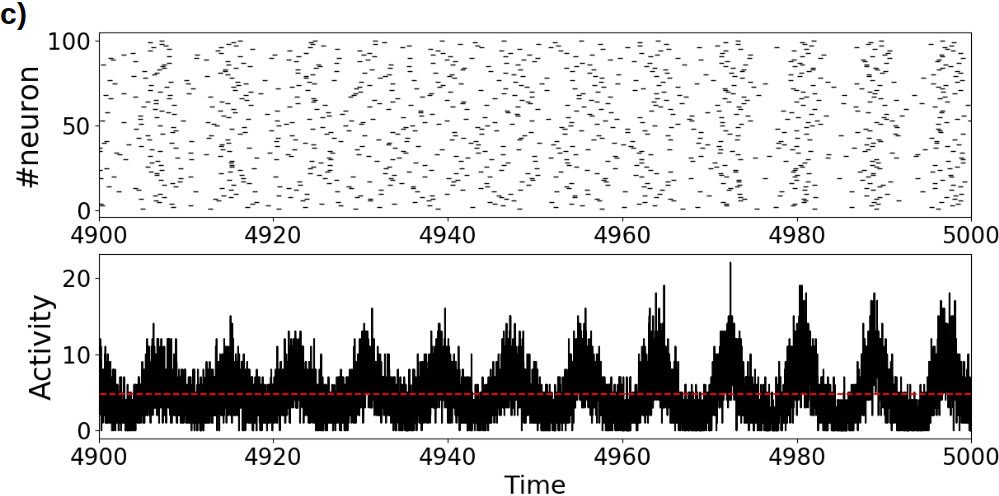}
   \includegraphics[width=0.48\textwidth]{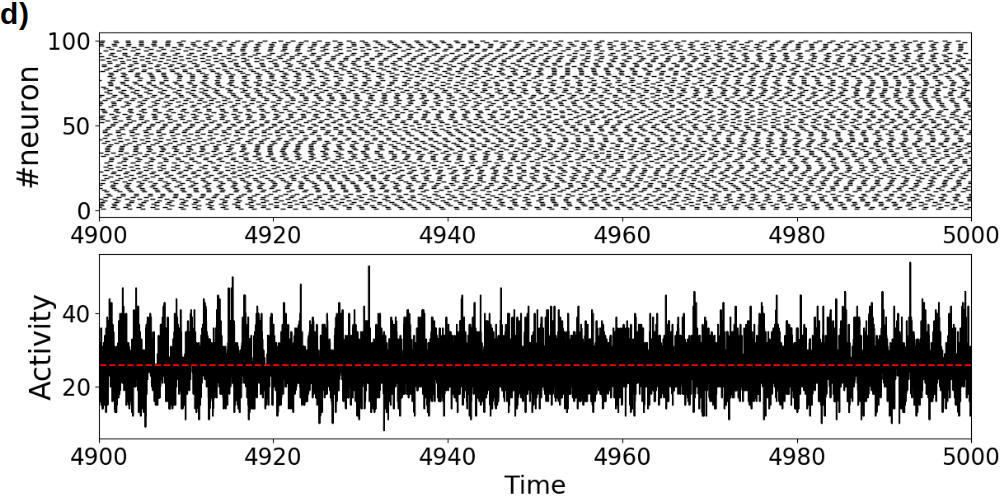}
   \includegraphics[width=0.48\textwidth]{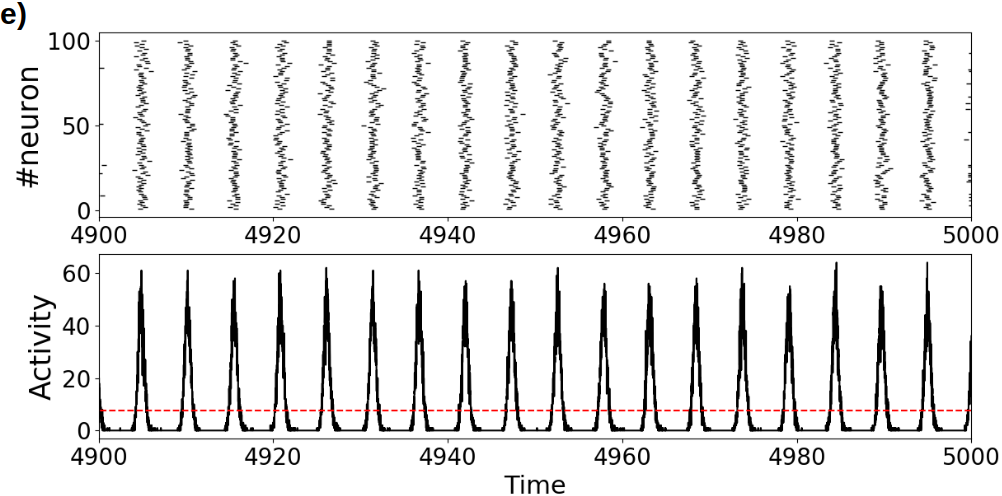}
\caption{The activity of the system for different points located at different parts of the phase diagram (Fig. \ref{phasediagram} (c)).For each point, the raster plot and the activity of the system are sketched. Also, the average of the activity of the network is shown by a red dashed line. The order parameter for $a=(g,\nu_{ext})=(1,2.5)$, $b=(2,1.5)$, $c=(8,1.17)$, $d=(4,3.5)$ and $e=(9,1.5)$ are $0.62,\, 0.001,\, 0.12,\, 0.001$ and $0.66$ respectively, which are completely consistent with their correspondent activity patterns. Point c is right at the margin of synchronous and asynchronous states where the system shows power-law behaviour. It has been investigated elaborately in the next part in terms of criticality.       }
\label{activity plots}
\end{figure}

 To see how the order parameter actually distinguishes the synchronous activity from an asynchronous activity, we have plotted the activity pattern of the system for a few points in the phase graph of Fig. \ref{phasediagram}(c) that is for $D=1.5$. The points selected are  $a=(g,\nu_{ext})=(1,2.5)$, $e=(9,1.5)$ and $d=(4,3.5)$ which lay in the L-Synch. region, R-Synch. region and the region where $m\simeq 0$, respectively. Actually, the order parameters of these points are  $m_a=0.62$, $m_e=0.66$ and $m_d=0.001$. Fig. \ref{activity plots} illustrates activity patterns with their corresponding raster plots of the system at these points. It is clear that the order parameter $m$ is completely sensitive to synchronization. We have checked similar graphs for many different points and found the order parameter consistent with the activity patterns.

\begin{figure}
\centering
    \includegraphics[width=7cm]{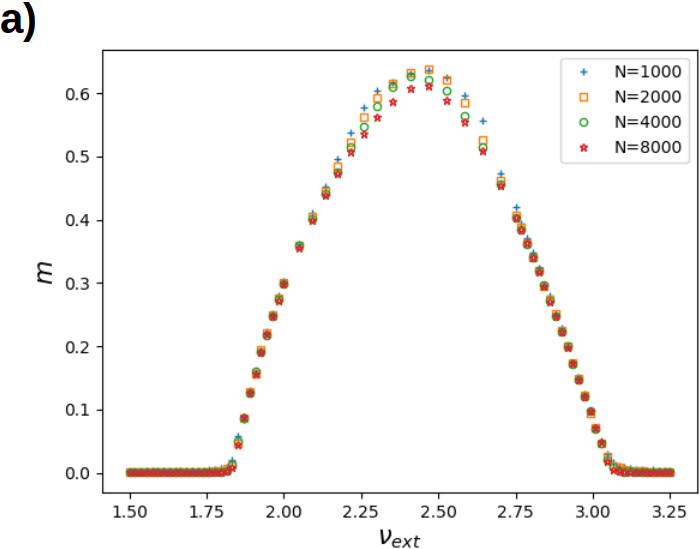}
    \includegraphics[width=7cm]{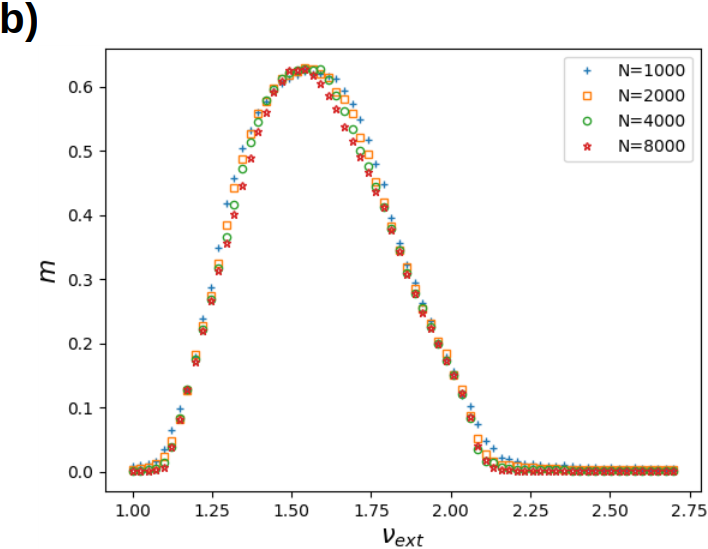}
    \includegraphics[width=7cm]{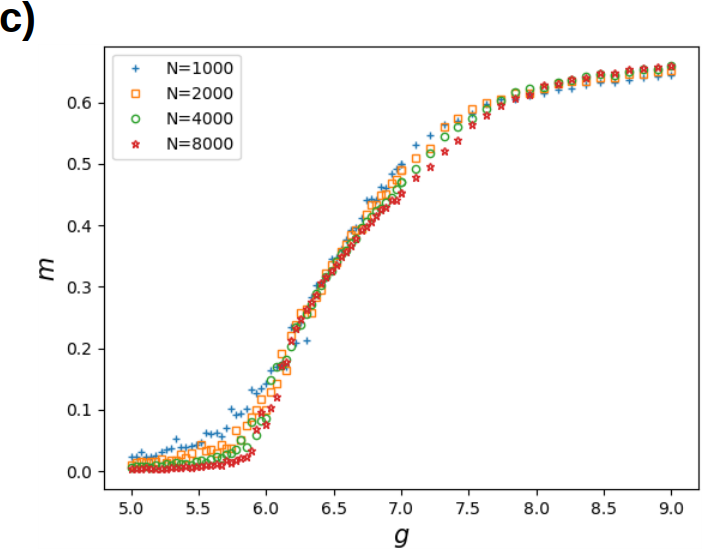}
  \caption {Three cross-sections of the order parameter for systems with $D=1.5$ and different sizes (the total phase diagram is shown for the system of 4000 neurons and various delays in Fig. \ref{phasediagram} ) .  a) The $g=1$ cross-section. b) The $g=1$ cross-section. c) The $\nu_{\rm ext}=1.5$. The cross-sections and the phase diagram approximately remains unchanged as we change the size of the system.    }
\label{phase diagram sections}
\end{figure}

The next point to be explored is if the synchronous regions are changed if the system size is altered. We have done simulations for sizes $N=1000, 2000, 4000$ and $8000$. Fig. \ref{phase diagram sections} shows a cross-section of the phase diagram at $g=8$, $g=1$ and $\nu_{ext}=1.5$. It is observed that the order parameter shows little dependence on the size of the system, the only difference is that the fluctuations are stronger in smaller systems. Also, as it is clearly seen in Fig. \ref{phase diagram sections}(c), larger networks have sharper transitions and in the limit $N\rightarrow \infty$ the derivative of the order parameter will become discontinuous which is a hallmark of second-order phase transition. However, as we are dealing with systems with sizes up to $N=16000$, we have to be careful about the finite size effects.

One last point that is worth mentioning: the order parameter does depend on the number of synapses per neuron, but as we have kept this quantity constant in all simulations shown here, this effect is not observed. This dependence can be expected, for example, if we reduce this number to zero there will be no synchronization and on the other hand, if all the neurons are connected to each other the synchronization happens much more easily.

\section{Results: Synchronization and power-law behaviour}
 
The phase diagram of the system (Fig.\ref{phasediagram}) suggests something like a continuous phase transition between synchronous and asynchronous regions. To assess this transition and identify the possible critical behaviour, we investigate neural avalanches in the network. In the literature, the existence of scale-free neural avalanches is the most important indicator of criticality in neural systems\cite{Beggs}. Generally, the network displays successive activities of various sizes $s$ and duration $d$, which are called avalanches. If the system is critical, the size and duration of avalanche distributions show power-law behaviour, i.e, $p(s)\sim s^{-\gamma_s}$ and $p(d)\sim d^{-\gamma_d}$\cite{Beggs}. 

\begin{figure*}[ht]
\centering
\begin{tabular}{cc}
    \includegraphics[width=0.47\linewidth]{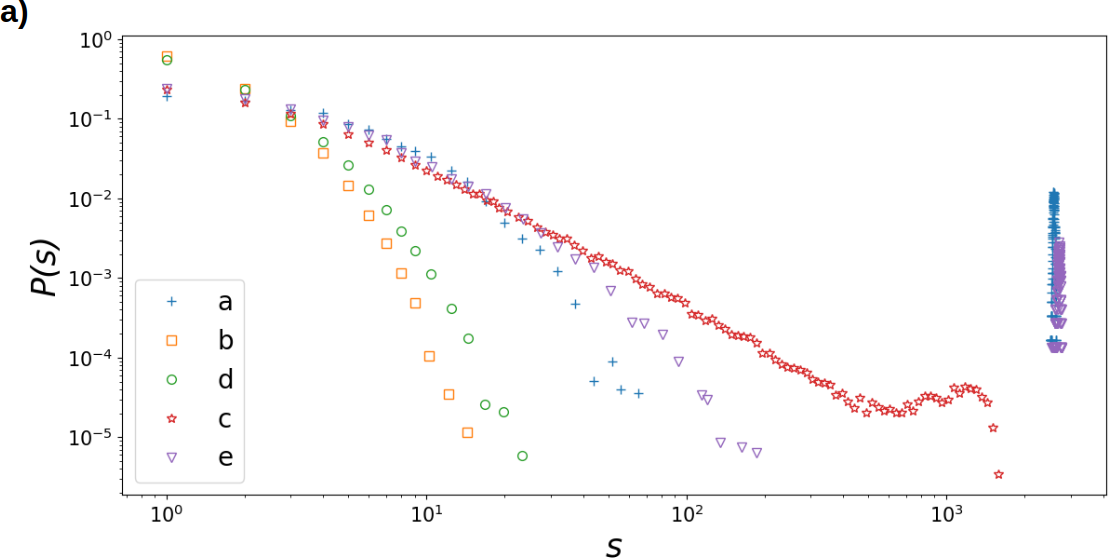}&
    \includegraphics[width=0.47\linewidth]{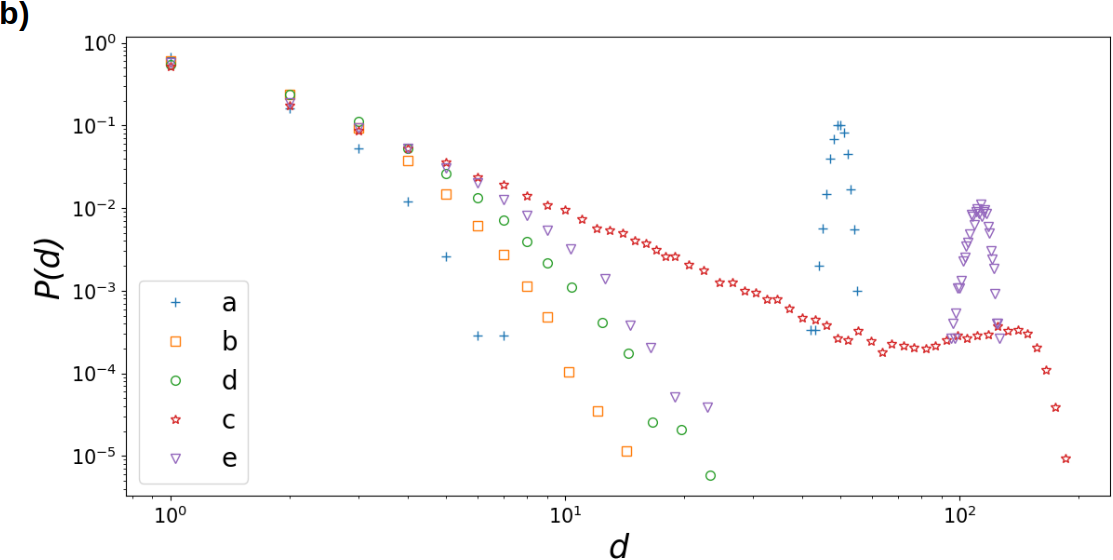}

\end{tabular}
\caption{Avalanche distributions for both size (a) and period (b) for some points of the phase diagram of the system shown in Fig. \ref{avalanche-general} (c). For sub-critical points (b,c) distributions in both plots illustrate Poissonian behavior. For supper-critical points (a,e), in addition to the Poissonian behavior, there is a bump at the end of the distributions due to the synchronization. Right between these two regimes, point c performs power-law distribution which is the indicator of criticality. This suggests a second-order phase transition between synchronous and asynchronous stats at this point. }
\label{avalanche-general}
\end{figure*}

Identification of successive avalanches is a challenging task. One may think of non-stop continuous activities within the system so that at least one quiescent temporal step $\Delta t$ exists between two separated avalanches. However, this definition depends on $\Delta t$, additionally for very large systems with lots of neurons, such a quiescent period may happen rarely. A better mechanism for identifying the avalanches is to consider the activity of the network and consider each period of time within which the activity is higher than the time-average of the network's activity as an avalanche. This definition has been used before in some similar systems \cite{Delpapa,Montakhab}. To avoid incorrect statistics, we have considered the activity over the threshold as the size of activity \cite{VeligasSanto}
Probability distributions of avalanche size and avalanche period, for networks of 4000 neurons and various values of $g$ and $\nu_{ext}$, are performed in Fig \ref{avalanche-general}.  Asynchronous systems, whose order parameter is nearly zero, show sub-critical behaviour with a Poisson like distribution, whereas synchronous systems that lie within  L-Synch. or R-Synch. regions,  display super-critical distributions. However, right at the transition curve, the probability distribution clearly shows a power-law distribution, though there is a bump in the end that usually is referred to as a sign of super-critical behavior. We will come back to this issue later. 


\begin{figure}
  \includegraphics[width=8cm]{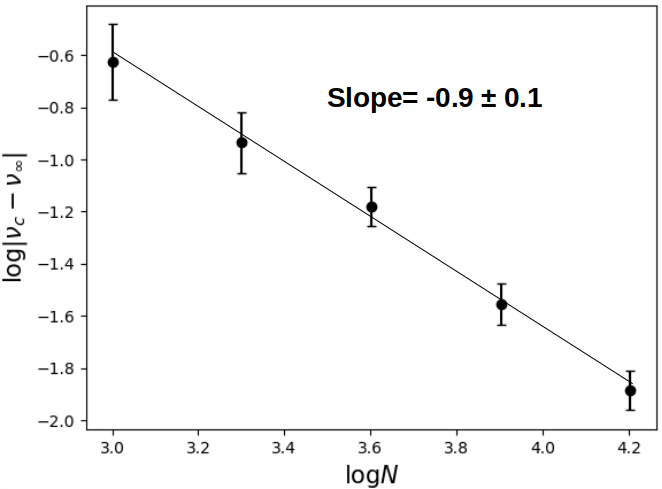}
  \caption{The log-log plot of $\nu_{\rm ext}$ at the critical point as a function of $N$, the number of neurons. The system shows a good scaling with the shifting exponent  $\lambda=0.9\pm0.1$. The best fit gives $\nu_\infty=1.098\pm0.005$ which is the critical $\nu_{ext}$ for an infinitely large system. }\label{Shifting}
\end{figure}

Consider a cross-section like the one shown in Fig. \ref{phase diagram sections}(a). Increasing $\nu_{\rm ext.}$ from unity, the shape of the probability distribution function changes from a Poisson-like curve to a power-law and then again to a Poisson-like curve but with a bump at very large values of $s$. Let us call the point that the system exhibits power-law behaviour by "the scaling point". This scaling point happens to be near the transition point of the system from the asynchronous region to the synchronous one.  Identification of the exact value of $\nu_{\rm ext.}$  for the scaling point in such a cross-section is hard because the probability distribution function gradually changes as we increase $\nu_{\rm ext.}$. In fact, moving from asynchronous region to the synchronous ones, a power-law behaviour is found without the bump in the end. Very quickly, with a little change in the external parameters, the bump forms and the probability distribution function keeps its form in a small interval of the external parameters and then the scaling behavior fades away. Although, usually in the literature of neural systems the presence of a bump is considered as a sign of super-criticality, this is not the case in the literature of critical systems and self-organized criticality. The presence of the bump can be a general feature in probability distribution functions to keep the total probability equal to unity. In addition to this reasoning, in our case, when the bump is present, due to the higher number of larger events we will have a finer tail with a less statistical error that allows us to check criticality through data collapse. Therefore, we consider the bumpy probability distribution to investigate the critical properties of the scaling point. We have also calculated indicators like $\langle s\rangle\langle s^3\rangle/(\langle s^2\rangle)^2 $ \cite{Pruessner} to identify the scaling point, however, such parameters show a relatively broad peak at the transition too, unless for our largest system sizes. For the cross-section $g=8$ and for systems with different sizes, the approximate values of $\nu_{\rm ext.}$ of the scaling point is brought in table \ref{tab:my_label}. For smaller systems, the scaling point is not so close to the transition point, but in the case of large systems the scaling point turns out to be right at the transition point.

\begin{table}[]
    \centering
    \begin{tabular}{|c||c|c|c|c|c|}\hline
         $N$ & 1000 &2000& 4000 &8000 &16000 \\ \hline
         $\nu_c (N)$& $1.34\pm0.09$&  $1.22\pm0.04$&$1.169\pm0.015$& $1.133\pm0.007$& $1.117\pm0.003$\\
         \hline
    \end{tabular}
    \caption{The critical point with its correspondent error for various sizes of the system. As the size of the system increases, the value of the error decreases. We have estimated $\nu_{\infty}=1.098\pm0.005$.  }
    \label{tab:my_label}
\end{table}
In fact, it can be shown that this is a finite-size effect. When the usual statistical mechanics' models are simulated, a size-dependent (pseudo)critical point is found. If we denote the (pseudo)critical point by $\nu_{C}(N)$ and the actual critical point by $\nu_\infty$ then we have
\begin{equation}
    |\nu_{C}(N)-\nu_\infty|\sim N^{-\lambda}
\end{equation}
where $\lambda$ is the shifting exponent. Fig. \ref{Shifting} shows a log-log plot of $\nu_{C}(N)-\nu_\infty$ as a function of size of the system $N$. The parameters $\nu_\infty$ and $\lambda$ are obtained by the best fitting. A very good linear fit is observed with $\nu_\infty=1.098\pm 0.005$ and $\lambda=0.9\pm 0.1$. The value obtained for $\nu_{\infty}$ is quite the same as the value of $\nu_{\rm ext.}$ at which the transition to synchronization occurs within the precision we have. In fact, for the line $g=8$ we have $\nu_{\rm ext.}=1.098\pm0.005$ for the transition point. We have checked the same phenomenon for a few other cross-sections, some of them being horizontal and some on the R-Synch. region, and in all cases the scaling point coincides with the transition point. Also, the scaling exponents obtained in the following are the same within error-bars.

\begin{figure*}[t]
\centering
\begin{tabular}{cc}
    \includegraphics[width=0.47\linewidth]{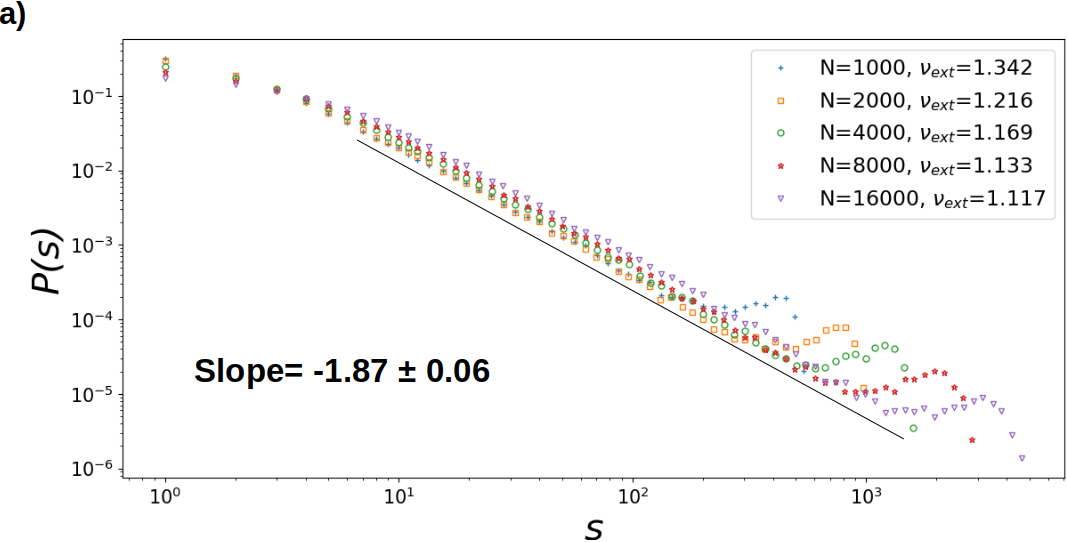}&
    \includegraphics[width=0.47\linewidth]{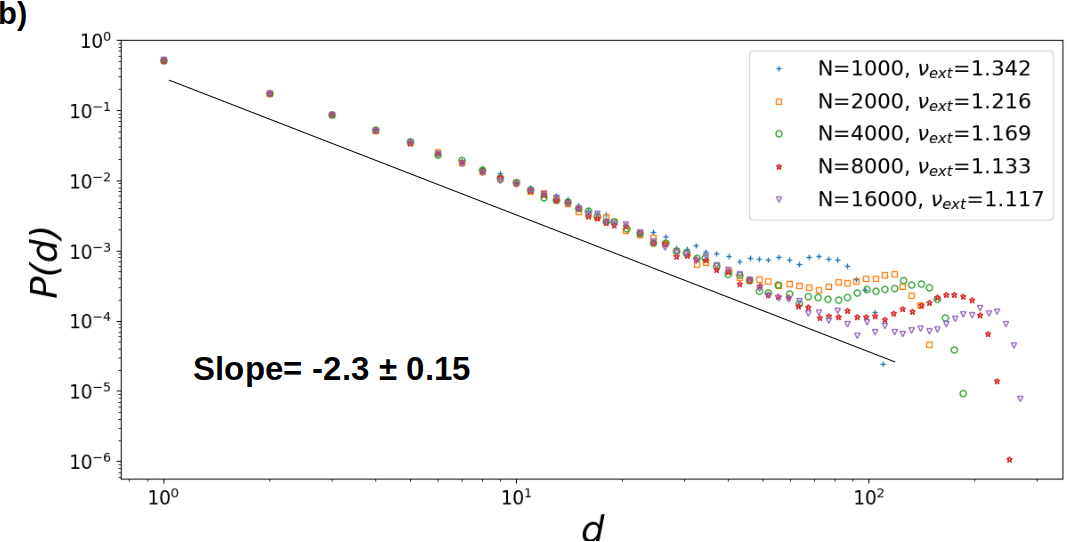}\\
    
\end{tabular}
\caption{Probability distribution function of size (a) and period (b) for systems of various sizes at their critical points. The exponents of the distributions are $\tau_s=1.87\pm0.06$ and $\tau_d= 2.3\pm0.15$ respectively for size and period.  }
\label{critical points}
\end{figure*}

Let us focus on the probability distribution function on the scaling point. Fig. \ref{critical points} shows the log-log plot of the probability distribution function of avalanche sizes and avalanche periods for different system sizes. Both plots show a linear part, which is the scaling region, and a cut-off that is clearly size-dependent, which is a hallmark of criticality. As said before, we have considered the external parameters where the probability distributions with a bump at the end of the plot. This consideration has no effect on the scaling exponents, however, gives a less statistical error because of finer data at the tail of the probability distribution. Note that the probability of avalanches at the bump is extremely small (for example $10^{-5}$ for system size $N=16000$ in the case of avalanche size distribution) and does not have an important role in the avalanche size statistics. The next interesting point is that in the case of avalanche size distribution, the linear part of the plot for systems with different sizes do not form a single line and for a system with a larger size, the line stands slightly above the line of a smaller system. This phenomenon, however, does not happen in the case of the period of avalanches. We will discuss it later and will see that the dependence causes some interesting phenomena to happen.

Despite this size dependency, the scaling exponent, $\tau_s$, does not depend on the size. We have obtained the exponents using the technique explained in \cite{Clauset}  and found $\tau_s=1.87\pm0.06$. Also for the case of the period of avalanches, we have $\tau_d= 2.3\pm0.15$. The values obtained from brain activity data are $1.5$ and $2$, respectively \cite{Beggs}. Considering the error bars of the two exponents obtained here, the exponent associated with the period of avalanches, $\tau_d$, is not much far from the experimental value, however, $\tau_s$ is completely different. As we show in the following, this exponent is an apparent exponent and finite-size scaling reveals another (real) exponent for the system.

To study the scaling behavior of the system more thoroughly, we have performed a Finite Size Scaling analysis (FSS) for both the avalanche sizes and their periods. If a system is critical and hence scale-invariant, the probability distribution function should be of the form 
\begin{equation}\label{Gen PDF}
    p(x)=x^{-\tilde{\tau}}f(x/N^\phi)
\end{equation}
where $x$ can be avalanche size or its period and  $\tilde{\tau}$ and $\phi$ are the scaling exponents. Of course, the above equation should be valid for $x>x_0$ where $x_0$ is a lower cut-off. The function $f$ is usually called the scaling function and depends only on the quantity $x/N^\phi$. 

\begin{figure*}[t]
\centering
\begin{tabular}{cc}
    \includegraphics[width=0.47\linewidth]{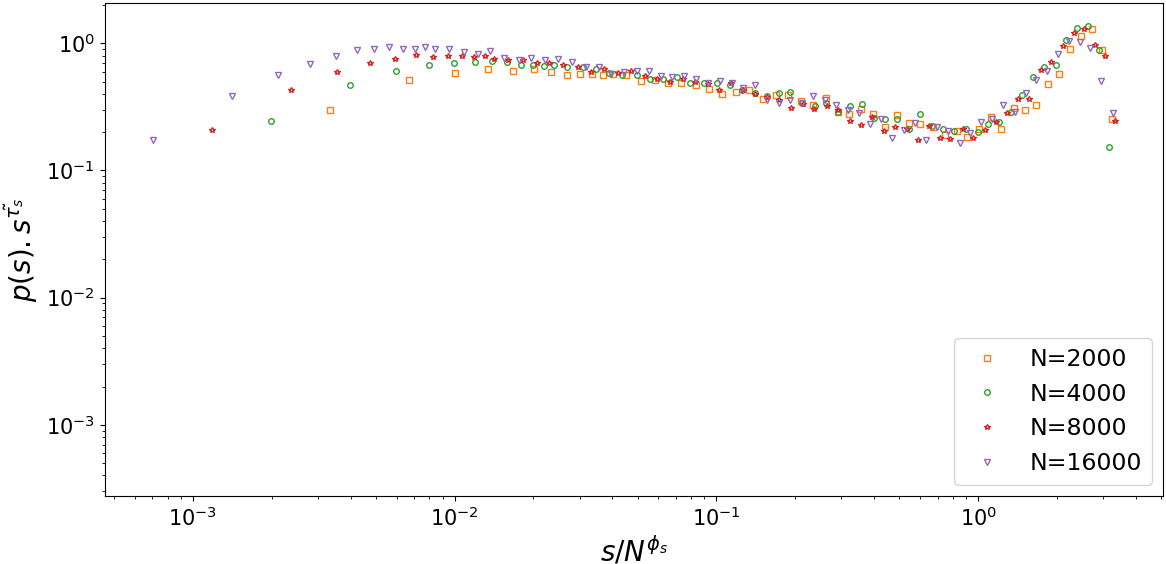}&
    \includegraphics[width=0.47\linewidth]{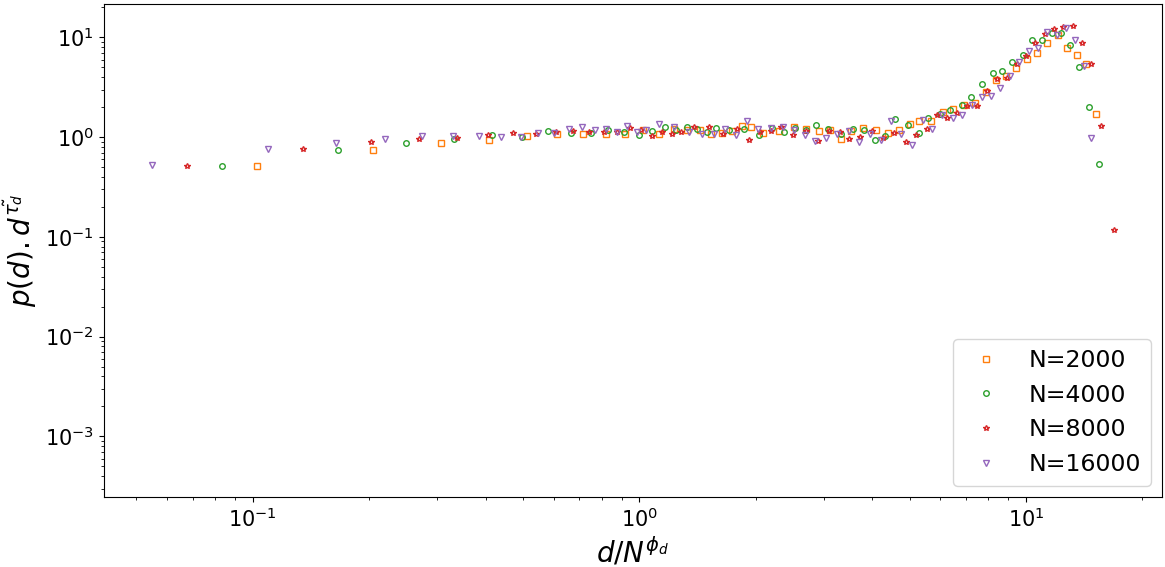}\\
\end{tabular}
\caption{Finite-size-scaling for a) avalanche size and b) avalanche period probability distribution function shown in Fig.\ref{critical points}. The scaling parameters are $\tilde{\tau}_s=1.53\pm 0.05$ and $\phi_s=0.72\pm 0.04$ for avalanche size and $\tilde{\tau}_d=2.3\pm 0.15$ and $\phi_d=0.33\pm 0.05$ for avalanche period.  }
\label{fss}
\end{figure*}

To check if a system obeys the above equation, we have to do a data collapse. This can be achieved by plotting  $x^{\tilde{\tau}}\times p(x)$ against $x/N^\phi$ for the different system sizes. If for all $x$ greater than a lower cutoff, the data of different system sizes $N$ can be collapsed onto a single curve, then it is manifestly shown that the probability distribution function has the scaling property. Fig. \ref{fss} shows such a data collapse for avalanche size and avalanche period. The scaling exponents are obtained to be $\tilde{\tau}_s=1.53\pm 0.05$ and $\phi_s=0.72\pm 0.04$ for avalanche size and $\tilde{\tau}_d=2.3\pm 0.15$ and $\phi_d=0.33\pm 0.05$ for avalanche period. Interestingly, the value obtained for the scaling exponent $\tilde{\tau}_s$ is different from $\tau_s$. Such a phenomenon has been observed before \cite{Chris-Prus}. In fact, it is usually assumed that the scaling function $f(u)$ has a finite value in the limit $u\rightarrow 0$. However, in some cases, the scaling function behaves as $x^\alpha$ in the limit mentioned above. If this is the case, we can write $f(x)=x^{\alpha}g(x)$ where $\lim_{x\rightarrow 0} g(x)=g_0$ is a finite number. Therefore, we will have 
\begin{equation}
    p(s)=s^{-\tilde{\tau}_s} \left(s/N^{\phi_s}\right)^\alpha g(s/N^\phi)=s^{-\tilde{\tau}_s+\alpha} N^{-\alpha \phi_s}g(s/N^\phi).
\end{equation}
 In the limit of large system sizes, for avalanche sizes much smaller than $N$, the quantity $s/N^\phi$ becomes very small and therefore, we can effectively take the argument of $g$ equal to zero. Then, the probability distribution function will be proportional $s^{-\tilde{\tau}_s+\alpha}$ and hence, the apparent scaling exponent $\tau=\tilde{\tau}_s-\alpha$ will be observed. As it is seen in Fig. \ref{fss}, the scaling function $f(u)$ does not have a finite value when its argument tends to zero, rather, it behaves as $u^{\alpha}$ with $\alpha=-0.33\pm 0.03$, that is, the scaling function becomes very large when $u\rightarrow 0$. Note that for each system size, there is a lower cut-off equal to $s_0/N^\phi$ below which the scaling function deviates from $g(u)\sim u^\alpha$. This analysis gives the apparent scaling exponent $\tau\simeq 1.86$ which was obtained before by fitting in Fig. \ref{critical points}.

\begin{figure}
  \includegraphics[width=8cm]{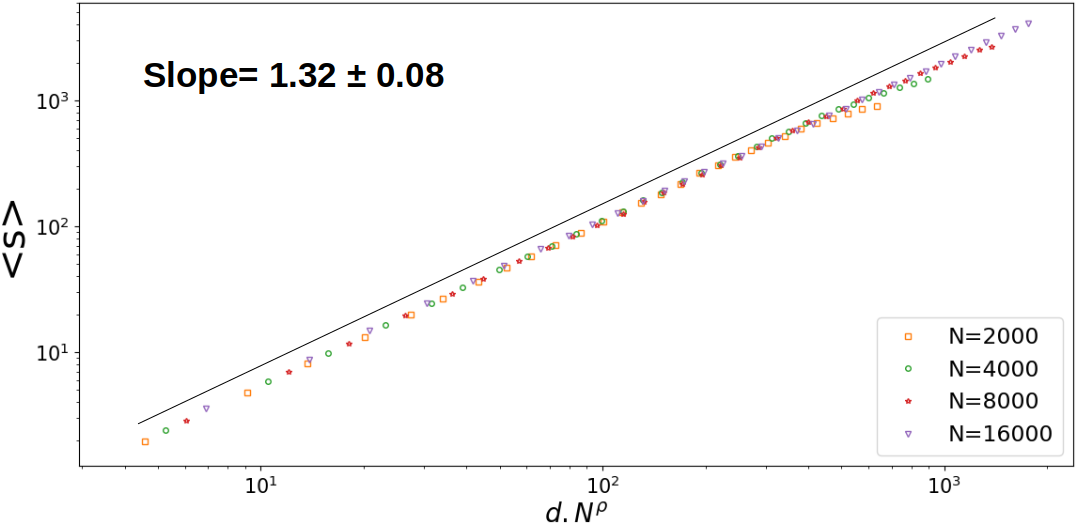}
  \caption{Average avalanche size versus period scaled by $N^\rho$. We obtain $\rho=0.2$ for data collapse $\gamma=1.32\pm 0.08$ for the best fit. }\label{average size versus period}
\end{figure}

 This size dependency of probability distribution function has some other consequences. As an example, consider the relation between mean avalanche size and mean avalanche period. Assuming narrow joint distributions,  $s$ and $d$ are functions of each other and we may write
 \begin{equation}\label{Joint Prob}
     \langle s\rangle_d \simeq d^{\gamma},
 \end{equation}
 where $ \langle s\rangle_d$ is the average of $s$ conditioned to $d$. In fact, Eq. \ref{Joint Prob} comes from two assumptions. First, the assumption  of narrow joint distribution and second, the fact that both the probability distribution functions of $s$ and $d$ are given by Eq. \ref{Gen PDF}. also, one may find a relation among the scaling exponents $\tilde{\tau}_s$, $\tilde{\tau}_d$ and $\gamma$. This relation is read to be:
 \begin{equation}
     \gamma=\frac{1-\tilde{\tau}_d}{1-\tilde{\tau}_s}.
 \end{equation}
 However, in this derivation it is assumed a finite value for $\lim_{u\rightarrow0} f(u)$. With a simple algebra, it can be shown that if $lim_{u\rightarrow0} f(u)\sim u^\alpha$, the relation between mean values of avalanche size and avalanche period can be written as
 \begin{equation}
     \langle s\rangle_d \simeq N^{\rho}d^{\gamma}
 \end{equation}
 with
 \begin{equation}
     \gamma=\frac{1-\tilde{\tau}_d}{1-\tilde{\tau}_s+\alpha}=\frac{1-\tilde{\tau}_d}{1-\tau_s}
 \end{equation}
 and
 \begin{equation}
     \rho=\frac{\phi_s\alpha}{1-\tau_s}.
 \end{equation}
 With the obtained values for scaling exponents we find $\gamma=1.5\pm 0.2$ and $\rho=0.24\pm 0.06$. Fig. \ref{average size versus period} shows a log-log plot of $\langle s\rangle_d$ for different systems sizes as a function of $d$ scaled by $N^\rho$. In this graph, we have assumed  $\rho=0.2$ and the best fit for the collapsed curves give $\gamma=1.32\pm 0.08$ which is consistent with some previously obtained value and experimental data\cite{Friedman 2012}, although originated from different scaling exponents $\tau$ and $\alpha$.





\section{Summary and conclusion}

In this paper, we studied the behavior of neural networks consisting of over-damped rotators as a model for the dynamic of type I single neurons.  Tuning external stimulation, inhibitory strength, and axonal delay time, we have identified both synchronous and asynchronous regions in the phase diagram of our system. Through finite size analysis, we have proposed that the transition is actually a  second-order phase transition between synchronous and asynchronous states in the system. Interestingly we have observed that the probability distribution function shows the power-law distribution for both size and period of avalanches in the vicinity of critical lines. At these lines, the probability distribution functions show a very good data collapse. The interesting point is that for the smaller systems, the point where the power-law behavior emerges falls inside the synchronous region. Therefore, in these systems, it is possible to observe both criticality and synchronization. Although,  it is seen that for a very large system, the phase transition happens right at the synchronization transition point.     

Yet there remain some unanswered questions. First of all, the transition line hardly touches the line on which the balance between inhibitory and excitatory neurons happens. Also, we have not proposed a mechanism that takes the system towards the critical line. One of the answers might be the evolution of synapses which is absent in our theory.  For example, if we devise a mechanism that strengthens the excitatory neurons at the synchronous phase and strengthens the inhibitory ones in asynchronous phase, the transition line to the r-synch region would become stable. However, at this point, we have not implemented any kind of synapse dynamics in the system.


\bibliography{brainsoc.bib}

\end{document}